\documentstyle[12pt,epsfig,rotating]{article}
\newcommand{\AmS}{{\protect\the\textfont2
  A\kern-.1667em\lower.5ex\hbox{M}\kern-.125emS}}
  \def\La{\Lambda} 
    
\def\E{\mbox{e}^+\mbox{e}^-}

\def\ifmath#1{\relax\ifmmode #1\else $#1$\fi}%

\def\s2{\hskip2pt}  \def\d{{\rm d}}    
\def\f{\left}   \def\g{\right}
\newcommand{\beqa}{\begin{eqnarray}} \newcommand{\eeqa}{\end{eqnarray}  }
\newcommand{\beqan}{\begin{eqnarray*}} \newcommand{\eeqan}{\end{eqnarray*}}
\newcommand{\beq}{\begin{equation}} \newcommand{\eeq}{\end{equation}  }
\def\cl{\centerline} \def\bcc{\begin {center}} \def\ecc{\end {center}}
\def\btbl{\begin{tabular}} \def\etbl{\end{tabular}}

\def\E877{{\footnotesize E877}}

\begin{document}

\null{}\vskip -1.2cm
\hskip12cm{\bf HZPP-0005}
\vskip-0.2cm

\hskip12cm July 5, 2000

\vskip1cm

\centerline{\Large  Thermal Freeze-out and Longitudinally Non-uniform}
\vskip0.3cm
 \centerline{\Large  Collective Expansion Flow}
\vskip0.3cm
\centerline{\Large in Relativistic Heavy Ion Collisions\footnote{Work
supported in part by the NSFC under project 19775018.}}
\vskip1cm
\centerline{\large FENG Shengqin, LIU Feng and LIU Lianshou}
\vskip0.4cm
\centerline{\small Institute of Particle Physics, Huazhong Normal University, 
Wuhan, 430079, China}

\vskip0.9cm

\vskip2cm
\begin{center}{\large ABSTRACT}\end{center}
\vskip0.5cm                           
\begin{center}\begin{minipage}{124mm}
{\small \hskip0.8cm
The non-uniform longitudinal flow model (NUFM) propsed recently is
extended to include also transverse flow. The resulting longitudinally
non-uniform collective expansion model (NUCEM) is applied to the 
calculation of rapidity distribution of kaons, lambdas and protons in 
relativistic heavy ion collisions at CERN-SPS energies. The model results 
are compared with the 200 A GeV/$c$ S-S and 158 A GeV/$c$ Pb-Pb collision 
data. The central dips observed in experiments are reproduced in a natural 
way. }
\end{minipage}\end{center}
\vskip3cm
PACS number(s): {25.75.-q }
\newpage
\noindent
{\bf I. Introduction}
\vskip0.5cm
\noindent
Relativistic heavy-ion collision offers a unique opportunity to study the
hot and dense matter under controlled laboratory conditions~\cite{Bass}.
Hadronic spectra from these reactions reflect the dynamic of the
hot and dense zone formed in the collision. The baryon density, established
early in the reaction, is an important factor governing the evolution of the
system. Comparison of model predictions with measured rapidity and transverse
momentum distributions constrains the possible dynamical scenarios of the
reaction, such as those for longitudinal and transverse flow. In addition,
the mechanism by which the incoming nucleons lose momentum during collision 
(baryon stopping) is an important theoretical problem~\cite{Capella}.

The rich physics of longitudinal and transverse flows is due to their
sensitivity to the system evolution at early time. 
Due to the high pressure of the system created during a heavy-ion collision, 
particles might be boosted in the transverse and longitudinal directions.  
The expansion and cooling of the heated and highly compressed matter could 
lead to a considerable collectivity in the final state.  The collective 
expansion implies space-momentum correlation in particle distributions at 
freeze-out. 

In this paper we present a study of the experimental results from the central 
collisions of Pb-Pb at 158 A GeV/$c$ and S-S at 200 A GeV/$c$ measured by 
NA49 and NA35 at CERN SPS~\cite{Alber}\cite{Appel}. The primary goal of 
the study is to provide a chance to compare the stopping power and radial 
symmetrical flow in collision systems of different sizes. Note that at SPS 
energies the duration time of the evolving system will be longer 
in comparison with that at the AGS energies, 
and it is even more reasonable to assume that a thermalized system expands 
collectively both in the longitudinal and at the same time in the
transverse direction.   

A big challenge for thermal models coming from experimental data is the 
central dips
in the rapidity distributions of heavier particles observed in the central 
collisions of nuclei both at AGS and at SPS energies. It is well known that
models with uniform longitudinal flow~\cite{Schned} cannot account for such 
dips.  In order to reproduce the experimentally observed 
central dips, a parameterization of  baryon chemical potential $\mu_{i}$ 
as function of flow rapidity $\eta$ is introduced in Ref.~\cite{Sollf}. 
However, the physical meaning of such a parametrization is unclear. 

In the present paper we will construct a {\em longitudinally non-uniform 
collective expansion model} (NUCEM) basing on the physical argument that 
the fireballs 
produced in the nuclear collision will keep some memory on the motion of the 
incident nuclei. The non-uniform longitudinal flow will be parametrized using
a geometrical picture proposed in Ref.~\cite{FLL}. The results of model
calculation are compared
with the experimental data on the rapidity distributions of baryons and strange 
particles in 200 A GeV/$c$ S-S and 158 A GeV/$c$ Pb-Pb collisions. 

The longitudinally non-uniform collective expansion model will be formulated 
in section II. The results of model calculation are presented and compared 
with the experimental data in section III.  A short summary and conclusion 
will be given in section IV. 

\vskip0.8cm
\noindent
{\bf II. Longitudinally non-uniform collective expansion flow } 
\vskip0.5cm
\noindent
In the non-unifrom longitudinal flow model (NUFM) proposed in 
Ref~\cite{FLL} only longitudinal flow was taken into account. In fact, the one
dimensional expansion presumably dominates initially because of
anisotropic initial conditions. But it is inconsistent to assume that
a thermalized system expands collectively only in longitudinal direction 
without 
generating transverse flow from the high pressure in the hydrodynamic
system. In case of central collisions the hydrodynamic flow should be computed 
in at least (2+1) dimensions if azimuthal symmetry is maintained.
Beside these theoretical necessities for the inclusion of transverse 
flow~\cite{Werner}, there are also evidence for the existence of this flow 
from the experimental phenomena~\cite{Gutbrod}\cite{Abbott}.

For the description of particle production in the collective expansion
of fire-ball we start from the 
formalism of Cooper and Frye~\cite{Cooper} which describes the single-particle
spectrum as an integral over a freeze-out hypersurface, thus summing the 
contributions from all space-time points at which the particles decouple from
the fireball:
\beq   
E\frac {\d^{3} n}{\d^{3} p}=\frac {g}{(2\pi)^{3}}
\int_{\sigma_{\rm f}} f(x,p)p^{\nu}\d^{3}\sigma_{\nu} ,
\eeq
\noindent
where $g$ is the degeneracy factor and $f(x,p)$ the momentum distribution at
space-time point $x$. In thermal models one takes $f(x,p)$ as a thermal 
equilibrium distribution and determines $\sigma_{\rm f}$ by a freeze-out
criterion for thermal decoupling~\cite{Schned}. At freeze-out, the Boltzmann 
approximation is sufficient, but we allow for a space-time dependence of 
the temperature $T$, the chemical potential $\mu$, and the flow velocity
$u^{\mu}$:
\beq  
f(x,p)=\exp{\f(-\frac {p \cdot u(x) - \mu(x)}{T(x)}\g)} .
\eeq  
\noindent
Focusing on central collisions, we assume azimuthal symmetry of the spatial
geometry and momentum distributions. 

\def\ath{{\rm tanh^{-1}}}
The assumption of longitudinal Bjorken flow~\cite{Bj} suggests to use 
longitudinal proper time $\tau=\sqrt{t^{2} - z^{2}}$
and space-time rapidity $\eta_{\rm l}=\ath(z/t)$ as suitable variables in
the $t$-$z$ plane. The transverse radial coordinate is denoted by $r$. In 
an azimuthally symmetric geometry of this kind it is practical to decompose
the velocity field in the following way:
\beq  
u^{\nu} = (\cosh{\eta_{\rm t}}\cosh{\eta_{\rm l}}, 
{\bf e_{\rm r}}\sinh
{\eta_{\rm t}}, \cosh{\eta_{\rm t}}\sinh{\eta_{\rm r}}) .
\eeq
\noindent
Here ${\bf e_{\rm r}}$ is the 2-dimensional unit vector in the radial
direction, $\eta_{\rm l} = \eta_{\rm l}(t,r,z)$ is the longitudinal 
flow rapidity by which each volume
element on the z-axis moves relative to the 
center of mass, and $\eta_{\rm t} = \eta_{\rm t}(t,r,z)$ is
the rapidity corresponding to the transverse flow of the volume element 
at position $(r,z)$ as seen from a reference point at 
$z$ on the beam axis moving with the local flow velocity there.

The momentum of the particle in the center-of-fireball system
can be parametrized as follows:
\beq 
u^{\nu}p_{\nu} = m_{\rm t}\cosh(y-\eta_{\rm l})
\cosh{\eta_{\rm t}}-p_{\rm t}\sinh{\eta_{\rm l}}\cos(\phi-\varphi)
\eeq
\noindent
Because of azimuthal symmetry in central collisions 
we can integrate over $\phi$ making use of
the modified Bessel function $I_{0}(\alpha) = (2\pi)^{-1}\int \limits_{0}^
{2\pi} e^{\alpha\cos{\phi}}\d\phi$. 

The geometry of the freeze-out hypersurface
$\sigma_{\rm f}$ is fixed as follows: In the time direction we take a surface
of constant proper time, $\tau=\tau_{\rm f}$. In $\eta_{\rm l}$ direction 
the freeze-out
volume extends only to a maximum space-time rapidity $\eta_{0}$, 
which is required by the finite available total energy and 
breaks longitudinal boost-invariance propsed by Bjorken.
In the transverse direction the boundary is given by 
$R_{\rm f}$, which describes a cylindrical fireball in the 
${\eta}$-$r$ space.

Having specified the freeze-out geometry and the distribution function at 
freeze-out we obtain the following thermal single-particle spectrum:
\beq  
\frac {\d^{2}N} {dydm^{2}_{\rm t}} = \frac {g}{2\pi}m_{\rm t}\tau_{\rm f}
{\int_{t_{\rm f}}} \d\eta_{\rm l} r \d re^{\mu/T}
e^{{-\tilde{\alpha}}\cosh(y-\eta_{\rm l})}\cosh{(y-\eta_{\rm l})}I_{0}(\alpha) ,
\eeq
\noindent
where $\tilde{\alpha} = (m_{\rm t}/T)\cosh\eta_{\rm t}$, 
$\alpha=(p_{\rm t}/T)\sinh{\eta_{\rm t}}$.
After integrating over
$m_{\rm t}$, we get the rapidity distribution as follows: 
\beqa    
\frac {dN}{dy}& =& \frac {g}{4\pi}\int 
\limits_{m_{\rm t}^{\rm lo}}^{m_{\rm t}^{\rm hi}}\d m_{\rm t}^{2} 
\int \limits_{-\eta_{0}}^{\eta_{0}}\d\eta_{l}
\int \limits_{0}^{R_{\rm f}}\d r\tau_{\rm f}(r)re^{\mu/T}
e^{{-\tilde{\alpha}}\cosh(y-\eta_{\rm l})} \nonumber \\
&\null{}&\hskip4cm \times m_{\rm t}\cosh(y-\eta_{\rm l})I_{0}(\alpha) .
\eeqa
\noindent
Here $m_{\rm t}^{\rm lo}$, $m_{\rm t}^{\rm hi}$ denote the experimental 
limits in which the spectrum was measured. The freeze-out radius $R_{\rm f}$
and the longitudinal extent of the fireball is fixed via the finite 
interval $(-\eta_{0},\eta_{0})$. $\eta_{\rm t}=\tanh^{-1}\beta_{\rm t}$ is 
the rapidity of transverse flow.  Replacing the transverse flow velocity 
profile by its radial average, we get
\beq 
\frac {dN}{dy}=\frac {g\tau_{\rm f}R_{\rm f}^{2}}{8\pi}
\int \limits_{m_{\rm t}^{\rm lo}}^{m_{\rm t}^{\rm hi}}
\d m_{\rm t}^{2}m_{\rm t}I_{0}(\alpha)
\int \limits_{-\eta_0}^{\eta_0}\d\eta\cosh(y-\eta_{\rm l})e^{\mu/T}
e^{{-\tilde{\alpha}}\cosh(y-\eta_{\rm l})}.
\eeq
\newpage
\begin{center}
\begin{picture}(250,450)
\put(-132,326)
{
{\epsfig{file=fig1.epsi,width=480pt,height=380pt}}
}
\put(-40,213)
{
{\epsfig{file=fig2.epsi,width=320pt,height=54pt}}
}
\end{picture}
\end{center}

\vskip-12.8cm 
\hskip3cm ($a$) \hskip7cm ($b$)
\vskip0.5cm

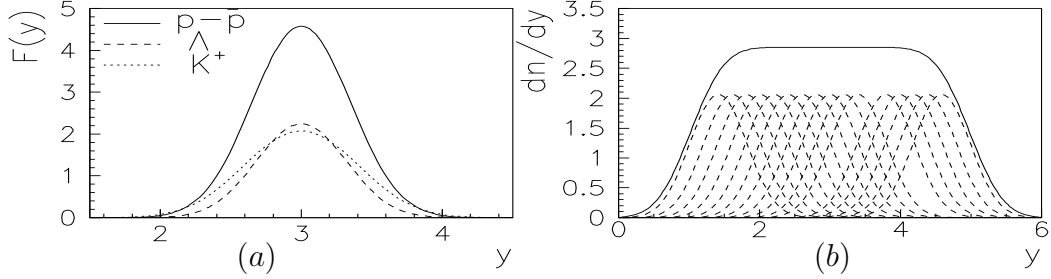
\cl{Fig.1($a$) \ Rapidity distribution from a single thermal fire-ball
 with transverse flow.}

\hskip0.8cm {($b$) \ Superposition of uniformly distributed kaon-fire-balls}

\vskip2.8cm 
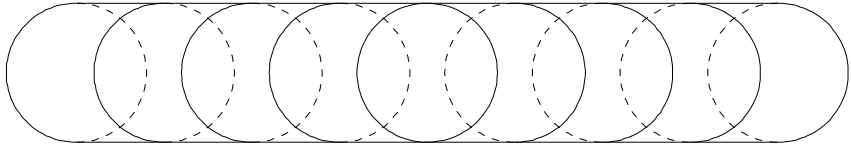
\cl{Fig.2 \ Shematic plot of the fire-balls distributed uniformly}

\hskip3cm {in longitudinal phase space}

In Eqn.(7) the collective expansions in both the longitudinal and the 
transverse directions have been taken into account. For the 
convenience of studying the non-uniform longitudinal flow, we change the 
order of integration in Eqn.(7) 
\beq 
\frac {dN}{dy}= \int \limits_{-\eta_0}^{\eta_0}\d\eta F(y-\eta_{\rm l}),
\eeq
\beq 
F(y)=\frac {g\tau_{\rm f}R_{\rm f}^{2}}{8\pi}
\int \limits_{m_{\rm t}^{\rm lo}}^{m_{\rm t}^{\rm hi}}
I_{0}(\alpha) \cosh(y)e^{\mu/T} e^{{-\tilde{\alpha}}\cosh(y)}
m_{\rm t} \d m_{\rm t}^{2}.
\eeq
The function $F(y)$ plotted in Fig.1 
can be interpreted as the rapidity distribution
from a ``{\em single thermal fire-ball with transverse flow}'' and the
rapidity distribution ${\d N}/{\d y}$ of Eqn's. (7), (8) 
is the sum of the contributions
of a series of such fire-balls with centers distributed uniformly in the  
rapidity range [$-\eta_0, \eta_0$], cf. the shematic plot in Fig.2 and 
Fig.1($b$).

It is well known that Eqn.(7), or equivalently Eqn's.(8)(9), does not
reproduce the central dip in the rapidity distirbutions of heavier particles 
in central relativistic heavy ion collisions observed experimentaly 
already at the AGS energies and confirmed further in the experiments at SPS
energies. In order to account for this experimental finding Ref.\cite{Sollf} 
introduced a rapidity dependence of chemical potential $\mu=\mu(\eta)$.
However, the physical meaning of this dependence is not so clear.

In the following we will follow the reasoning proposed in Ref.\cite{FLL},
i.e. we assume that the fire-balls produced in relativistic heavy ion
collisions will keep some memory on the motion of the incident nuclei, 
and therefore the distribution of fire-balls, instead of being uniform 
in the longitudinal direction, is more concentrated in the direction of 
motion of the incident nuclei. This means that the distribution of fire-ball 
is more dense at large absolute value of rapidity, as sketched  schematically 
in Fig.3. 

\begin{center}
\begin{picture}(250,450)
\put(-40,380)
{
{\epsfig{file=fig3.epsi,width=320pt,height=54pt}}
}
\put(-10,90)
{
{\epsfig{file=fig4.epsi,width=260pt,height=254pt}}
}
\end{picture}
\end{center}

\vskip-13.5cm 
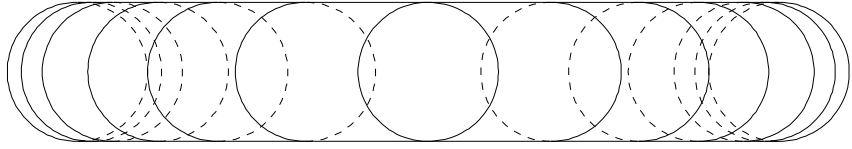
\cl{Fig.3 \ Schematic plot of the fire-ball distribution}

\hskip3cm in non-uniform longitudinal flow

\vskip8.5cm 
\cl{Fig.4\ The dependence of longitudinal distribution of fire-ball}

\hskip3cm {center $\rho(y_{\rm le})$ on ellipticity parameter $e$ }
\vskip0.5cm

Using a simple geometrical picture to parametrize the non-uniform longitudinal 
flow~\cite{FLL}, a non-flat distribution function of fire-ball center 
$y_{\rm le}$
\beq  
\rho(y_{\rm le}) = \sqrt {\frac {1 + \sinh^{2}(y_{\rm le})}
{1 + e^{2}\sinh^{2}(y_{\rm le})}} 
\eeq
is introduced and the rapidity distribution (8) becomes
\newpage

\begin{center}
\begin{picture}(250,450)
\put(-20,205)
{
{\epsfig{file=fig5.epsi,width=240pt,height=254pt}}
}
\end{picture}
\end{center}

\vskip-7.5cm 
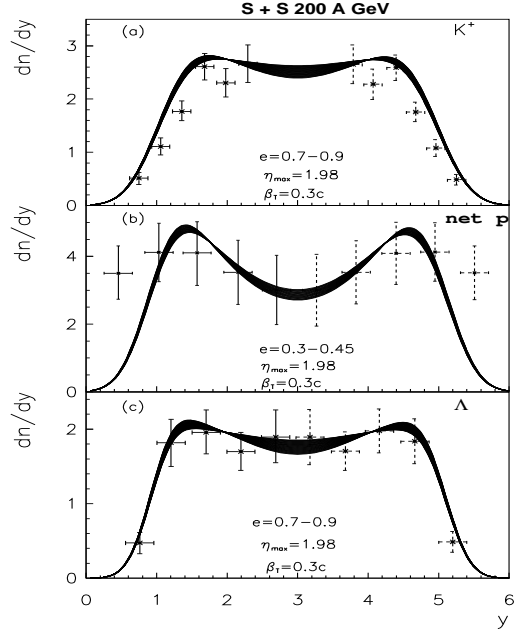
\cl{Fig.5\ Rapidity distributions of kaon ($a$), net proton ($b$) and 
lambda ($c$) }

\hskip1cm{ for central S-S collisions at 200 A GeV/$c$. Data are taken from 
Ref.\cite{Alber} }

\beqa 
\frac {dN}{dy} &=& \int \limits_{-y_{\rm le0}}^{y_{\rm le0}} \d y_{\rm le}
\rho(y_{\rm le}) F(y-y_{\rm le})\nonumber \\
&=& 
\frac {g\tau_{\rm f}R_{\rm f}^{2}}{8\pi}
\int \limits_{-y_{\rm le0}}^{y_{\rm le0}}
\rho(y_{\rm le})\d y_{\rm le}
\int \limits_{m_{\rm t}^{\rm lo}}^{m_{\rm t}^{\rm hi}}
\d m_{\rm t}^{2}m_{\rm t}I_{0}(\alpha) 
\cosh(y-y_{\rm le})e^{\mu/T}
e^{{-\tilde{\alpha}}\cosh(y-y_{\rm le})}
\eeqa

In Eqn.(10) $e$ is the ``ellipticity'' parameter describing the degree
of non-uniformity of longitudinal flow~\cite{FLL}.  The dependence of 
$\rho(y_{\rm le})$ on $e$ is shown in Fig.4.  It can be seen from the 
figure that the larger is the parameter $e$, the flatter is the distribution 
$\rho(y_{\rm le})$ and the more uniform is the longitudinal-flow distribution,
cf. Fig.3.  When $e\rightarrow{1}$, 
the longitudinal-flow distribution is completely uniform 
($\rho(y_{\rm le})\rightarrow{1}$), returning back to the 
cylindrically symmetrical collective flow model, Eqn.(7). 

\vskip0.6cm
\noindent
{\bf III. Comparison with experiments  } \\
\vskip0.2cm

We present in Fig.5 ($a, b$ and $c)$ the rapidity distributions of kaon, net 
proton and lambda for central S-S collisions~\cite{Alber} at 200 A GeV/$c$, 
respectively. In view of the large statistical errors of the experimental
data, the fitted values of parameter $e$ are given in a region and 
accordingly the rapidity distributions have a strip shape.

\begin{center}
\begin{picture}(250,450)
\put(-40,280)
{
{\epsfig{file=fig6.epsi,width=320pt,height=204pt}}
}
\end{picture}
\end{center}

\vskip-9.5cm
\cl{Fig.6\ The dependence of longitudinal distribution of fire-ball}

\hskip3cm {center $\rho(y_{\rm le})$ on ellipticity parameter $e$ }
\vskip0.5cm

The rapidity limit $y_{\rm le0}$ and the ellipticity $e$
used in the calculation are listed in Table I
and illustrated in Fig.6 (a). 
The parameter $T$ is chosen to be 0.12 GeV.

\vskip0.5cm
\cl{Table I \ \ The values of model-parameters}
\vskip-0.5cm
\bcc\btbl{|c|c|c|c|c|}\hline
           & \multicolumn{3}{|c|}{S-S Collisions}
        &\multicolumn{1}{|c|}{Pb-Pb Collisions} \\ \cline{2-5}
 Parameter & $k^{+}$ &p$-\bar{\rm p}$&$\La$ & p$-\bar{\rm p}$  
\\ \hline
$e$ & 0.7-0.9 & 0.3-0.45 & 0.7-0.9 & 0.56
\\ \hline
$y_{\rm le0}$ & 1.98 & 1.98 & 1.98 & 1.96  
\\ \hline
$<\beta_{t}>$ & 0.3c & 0.3c & 0.3c& 0.5c   \\
\hline
\etbl\ecc

Since kaons and lambdas are produced through the interaction of colliding
nuclei, they have less memory on the motion of the incident nuclei.
Therefore, the values of ellipticity $e$ for kaon and lambda
are bigger than that for proton, as shown in Table I.

It can be seen from the figures that our longitudinally non-uniform collective 
expansion model (NUCEM) reproduces the 
central dip of rapidity distributions for kaon, lambda and net proton
very well, in agreement with the experimental findings. 

The appearance
or disappearance of central dip is insensitive to the rapidity limit 
$y_{\rm le0}$ but depends strongly on the magnitude of the ellipticity 
$e$ and the mass $m$ of produced particles.   
When transverse flow exists there is a shallow dip also for
the rapidity distribution of light particle (kaon), cf. Fig.5, 
which is slightly different from the case~\cite{FLL} of one-dimensional 
longitudinal flow.

The rapidity distribution of net protons 
for Pb-Pb collisions at 158 A GeV/$c$~\cite{Appel} is shown in Fig.7,
and the longitudinal distributions of fire-ball center are showm in 
Fig.6($b$) for various values of ellipticity $e$. 
The solid line in Fig.7 corresponds to the results of our model 
with ellipticity parameters $e=0.56$. The fitted
$\chi^{2}/$DF is given in Fig.8. The value $e=0.56$ corresponds to
the minimum of $\chi^{2}$. The
region of $e$ for $\chi^{2}$ to increase one unit from the minimum
is between 0.40 and 0.745. 

\begin{center}
\begin{picture}(250,450)
\put(-60,300)
{
{\epsfig{file=fig7.epsi,width=200pt,height=264pt}}
}
\put(150,300)
{
{\epsfig{file=fig8.epsi,width=160pt,height=164pt}}
}
\end{picture}
\end{center}

\vskip-11.0cm 
\cl{Fig.7 \ Rapidity distribution of net protons 
\hskip1.5cm 
Fig.8\ The fitting $\chi^{2}$ }

\vskip0.5cm
Comparing Fig's.5 and 7 we can see that the net-proton
(p$-\bar{\rm p}$) rapidity distribution is narrower for Pb-Pb~\cite{Appel}, 
than for S-S collisions~\cite{Alber}. This indicates an increasing baryon 
stopping for Pb-Pb collisions, giving accordingly a smaller 
rapidity limit $y_{\rm le0}$ and a larger
average transverse velocity and uniformity for Pb-Pb interaction.
The width of the rapidity distributions is mainly
controlled by the amplitude $y_{\rm le0}$ of the longitudinal flow. 
For smaller colliding system (S-S collision) a single value of 
$y_{\rm le0}$ can account for the wide distribution
of heavier particles (net protons and $\La$) and at the same time fit the 
kaon-distribution well. For the larger colliding system (Pb-Pb),
the $y_{\rm le0}$  is smaller together with a larger 
average transverse flow velocity. These are shown also in Table I.  

\vskip1.2cm
\noindent
{\bf IV. Summary and Conclusions}
\vskip0.2cm
Let us begin with a discussion of the thermal freeze-out points. Freeze-out
marks the transition from a strongly coupled system, which evolves from
one state of local thermal equilibrium to another, to a weakly coupled
one of essentially free-streaming particles. If this transition happens 
quickly enough, the thermal momentum distributions (superimposed by 
collective expansion flow) are frozen in, and the temperature and 
collective  flow velocity at the transition point can be 
extracted from the measured momentum spectra. In high energy heavy-ion 
collisions the freeze-out process is triggered dynamically by the 
accelerating transverse expansion and the very rapid growth of the
mean free paths as a results of the fast dilution of the 
matter~\cite{Mayer}. Idealizing the kinetic freeze-out process by a single 
point in the phase diagram is therefore not an entirely unreasonable
procedure.

In high energy heavy-ion collisions, due to the transparency of nucleus 
the participants will not lose the historical memory totally 
and the produced hadrons will carry some of their 
parent's memory of motion, leading to the unequivalence in longitudinal
and transverse directions, i.e. the flow of 
produced particles is privileged in the longitudinal direction. 
As the increasing of duration time and the
pressure of the system at SPS energy, it is 
reasonable to assume that the system expands
collectively both in the transverse and in the longitudinal directions.
This picture has been used by lots of 
models~\cite{Schned},\cite{Braun}. For asymptotically high energies,  
a  boost-invariant longitudinal expansion model 
is postulated by Bjorken \cite{Bj}, which gives a plateau in the 
rapidity distribution of produced particles. For finite collision energies 
such as CERN SPS energy or below, a cylindrically symmetrical expansion
model was firstly postulated by Schnedermann, 
Sollfrank and Heinz\cite{Schned} by introducing a cut in rapidity. 
In this model a set of fire-balls with centers located uniformly in 
the rapidity 
region [-$y_{\rm le0}$, $y_{\rm le0}$], as sketched schematically in Fig.2,
represents the longitudinal flow, and at the same time there is 
radial flow developed simultaneously. 
It can account for the wider rapidity distribution
when compared with  the prediction of the pure thermal 
isotropic model but failed to reproduce
the central dip in the proton and $\La$ rapidity distributions.  

In this paper, we argue that the transparency/stopping of relativistic heavy 
ion collisions should be taken into account more carefully. It will not
only lead to the anisotropy in longitudinal-transverse directions, but 
also render the fire-balls 
to concentrate more in the direction of motion of the incident nuclei. 
A non-uniform longitudinal flow model is proposed, 
which assumes that the centers of fire-balls are distributed non-uniformly
in the longitudinal phase space.  A parameter $e$ is introduced through a
geometrical parameterization which can express the non-uniformity
of flow in the longitudinal direction, i.e.
the centers of fire-balls of produced particles 
prefer to accumulate in the two extreme
rapidity regions ($y_{\rm le} \approx \pm y_{\rm le0}$)
in the c.m.s. frame of relativistic heavy-ion collisions, 
and accordingly the distribution is diluted 
in the central rapidity region ($y_{\rm e} \approx 0$). 
Apart from the non-uniform longitudinal flow, a radial flow
in the transverse direction is also considered in the present paper.

It is found that the depth of the central dip  
depends on the magnitude of the parameter $e$ 
and the mass of produced particles, i.e. the 
non-uniformity of longitudinal flow
which is described by the parameter $e$ determines the depth of the central 
dip for produced particles. Comparing with one-dimensional non-uniform 
longitudinal flow model~\cite{FLL}, the  rapidity distribution of lighter
strange particle kaon also shows a dip due to the effect of 
transverse flow.  

Smaller colliding systems distinguish themselves from larger ones not
primarily by the achieved maximal energy density, but by the occupied
collision volume in space and time. Compared to S-S collisions, Pb-Pb
collisions live longer until thermal freeze-out, expand more in the 
transverse direction, develop more transverse collective flow.
It is found that transverse collective 
expansion is expected to be more prominent in the
heavy Pb-Pb collision system,
owing to smaller surface-to-volume ratio, the larger fireball volume and the 
longer duration time of expansion. On the  other hand,
Pb-Pb collisions develop less longitudinal flow, which suggests, 
together with the larger $e$,
a larger stopping in the bigger colliding system. 


\vskip2cm
\begin{thebibliography}{3}
\bibitem{Bass} S.A. Bass, M. Gyulassy, H. St$\ddot{o}$cker and W. Greiner,
J. Phys. {\bf G25}, (1999)R1-R57;
T. Alber et al., Phys. Rev. Lett. {\bf 75}, (1995) 3814.
\bibitem{Capella} A. Capella and B. Z. Kopeliovich, Phys. Lett. {\bf B381},
(1996)325; D. Kharzeev, Phys. Lett. {\bf B378}, (1996) 238; K. Werner,
Phys. Rep. {\bf 232},(1993)87. 
\bibitem{Letess} J. Letessier et al., Phys. Rev. {\bf D51}, (1995)3408.
\bibitem{Alber} E. Alber et al., (NA35 Collab.) Nucl. Phys. 
{\bf A566}; B$\ddot{a}$chler et al., (NA35 Collab.) Phys. Rev. Lett.
{\bf 72} (1994)1419. 
\bibitem{Appel} H. Appelsh$\ddot{a}$user et al.,(NA49 Collab) 
Phys. Rev. Lett. {\bf 80} (1998)4136.
\bibitem{Sollf}J. Sollfrank, Eur. Phys. J. {\bf C9}, (1999) 156;  C. Slotta, J
Sollfrank and U. Heinz, 1995 Strangeness in
Hadronic Matter, AIP Conference Proceedings {\bf 340} (Woodbury: AIP
Process) p 462. 
\bibitem{FLL} Shengqin Feng, Feng Liu and Lianshou Liu, {\em Thermal
equilibrium and non-uniform longitudinal flow in relativistic
heavy ion collisions}, hep-ph/0005047, to appear in Phys. Rev. C.
\bibitem{Werner} K. Werner, Quark Matter 1990, Menton, France, Nucl. Phys.
{\bf A525} (1991) 501c; 
K. Werner, P. Koch, Phys. Lett. {\bf B 242} (1990)251;  A.M. Poskanzer,
and S.A. Voloshin, Phys. Rev. {\bf C 58}, (1998) 1671;  J.-Y.
Ollitrault, Nucl. Phys. {\bf A 638}, (1998) 195c.
\bibitem{Gutbrod} H.H. Gutbrod, A.M. Poskanzer, H.G. Ritter, Rep. Prog. Phys.
{\bf 52} (1989) 1267;  H.A. Gustaffson et al., Phys. Rev. Lett. {\bf 53}
(1984) 1590;  H. St$\ddot{a}$cker, W. Greiner, Phys. Rep. {\bf 137} (1986)
277.
\bibitem{Abbott} T. Abbott et al., Phys. Rev. Lett. {\bf 64} (1990) 847;
J. Barrette et al., (E877 Collab) Phys. Rev. Lett. {\bf 73}, (1994) 2543.
\bibitem{Cooper} F. Cooper, G. Frye, Phys. Rev. {\bf D 10}, (1974) 186.
\bibitem{Schned} E. Schnedermann, J. Sollfrank and U. Heinz, Phys. Rev. 
{\bf C 47}, 1738(1993);  {\bf C 48}, (1993) 2462;
E. Schnedermann and U. Heinz, Phys. Rev. {\bf C 50}, (1994) 1675. 
\bibitem{Bj} J.D. Bjorken, Phy. Rev. {\bf D 27}, (1983) 140.
\bibitem{Braun} P. Braun-Munzinger, J. Stachel, J.P. Wessels and N. Xu,
Phys. Lett. {\bf B 344}, (1995) 43;  P. Braun-Munzinger, J. Stachel, 
J.P. Wessels and N. Xu, Phys. Lett. {\bf B 365}, (1996) 1; P. Braun-Munzinger, 
J. Stachel, Nucl. Phys. {\bf A 606}, (1996) 320.
\bibitem{Mayer} U. Mayer and U. Heinz, Phys. Rev. {\bf C 56}, (1997) 439;
E. Schnedermann and U. Heinz, Phys. Rev. {\bf C 50}, (1994) 1675;  U. Heinz,
Nucl. Phys. {\bf A 661}, (1999) 140c.
\end{thebibliography}
\end{document}